\documentclass[twocolumn]{aastex63}
\usepackage{CJK}
\usepackage{booktabs}
\newcommand{\ba}{\begin{eqnarray}}
\newcommand{\ea}{\end{eqnarray}}
\usepackage[normalem]{ulem}

\begin{document}
\begin{CJK*}{UTF8}{gbsn}
\title{Double Hot Jupiter Formation through Mirrored ZLK Migration in Binary Star Systems: \\ The Case of WASP-94}

\author[0009-0007-9211-2884]{Yurou Liu (刘雨柔)} 
\affiliation{Department of Astronomy, Yale University, New Haven, CT 06511, USA}

\author[0000-0003-0834-8645]{Tiger Lu (陆均)}
\altaffiliation{Flatiron Research Fellow}
\affiliation{Center for Computational Astrophysics, Flatiron Institute, 162 5th Avenue, New York, NY 10010, USA}
\affiliation{Department of Astronomy, Yale University, New Haven, CT 06511, USA}

\author[0000-0002-7670-670X]{Malena Rice (米乐娜)}
\affiliation{Department of Astronomy, Yale University, New Haven, CT 06511, USA}


\correspondingauthor{Yurou Liu}
\email{yurou.liu@yale.edu}

\begin{abstract}
To date, only a handful of binary star systems are known with at least one confirmed planet orbiting each star. Such systems, however, offer a unique perspective on the stochasticity intrinsic to planet formation and evolution -- particularly in twin binary star systems, which consist of near-equal-mass stars formed contemporaneously in the same birth environment. The WASP-94 system, which includes twin F-type stars, is a striking exemplar of such systems, containing \textit{two} hot Jupiters: WASP-94 Ab is a transiting, spin-orbit misaligned giant planet with a $4$-day orbital period, while WASP-94 Bb is non-transiting and has a tighter $2$-day orbital period. In this work, we leverage \textit{N}-body simulations to show that the current double hot Jupiter configuration of the WASP-94 system can be reproduced through mirrored von Zeipel-Lidov-Kozai migration. The upcoming \textit{Gaia} astrometric data releases offer the potential to search for additional twin planetary systems, including double cold Jupiter systems that may serve as the progenitors for WASP-94-like configurations.

\end{abstract}

\section{Introduction}

A foundational question underlying exoplanet demographics is that of how deterministic the outcomes of planet formation are -- namely, to what extent do similar initial conditions produce similar system outcomes?

Astronomy does not lend itself naturally to standard laboratory experiments with control samples, so this question cannot be tested directly. However, this limitation can be circumvented through characterizing populations of natural laboratories with built-in versions of ``control samples'': specifically, by examining comparative demographics in planet-hosting twin binary star systems, with near-equal-mass stars each born in similar environments \citep{hand2025case}. After their initial formation, such systems may undergo dynamical evolution that further shapes the 3D orbital geometries observed today.

One particularly intriguing such system is WASP-94: a twin binary system with one hot Jupiter found around each of the two F-type stars. WASP-94 is one of only a few known binary star systems in which both stars host one or more circumstellar planets, alongside HD 20782/20781 \citep{udry2019harps}, XO-2N/S \citep{desidera2014gaps}, HD 133131A/B \citep{Kai-Uwe2024gaiacompanion}, and the recently proposed candidate system TOI-2267A/B \citep{Zuniga-Fernandez2025TOI2267}.\footnote{We also note that HD 41004 A/B hosts at least one companion around each star. We exclude this system from the list because HD 41004 B b has $m\sin I=18.37\pm0.22M_{\rm Jup}$ \citep{zucker2004multi}, placing it in the brown dwarf mass regime and suggesting that it may have formed through a distinct channel.} Among these systems, WASP-94 is the only one in which each star hosts a hot Jupiter.

\begin{figure*}
    \centering
    \includegraphics[width=0.8\linewidth]{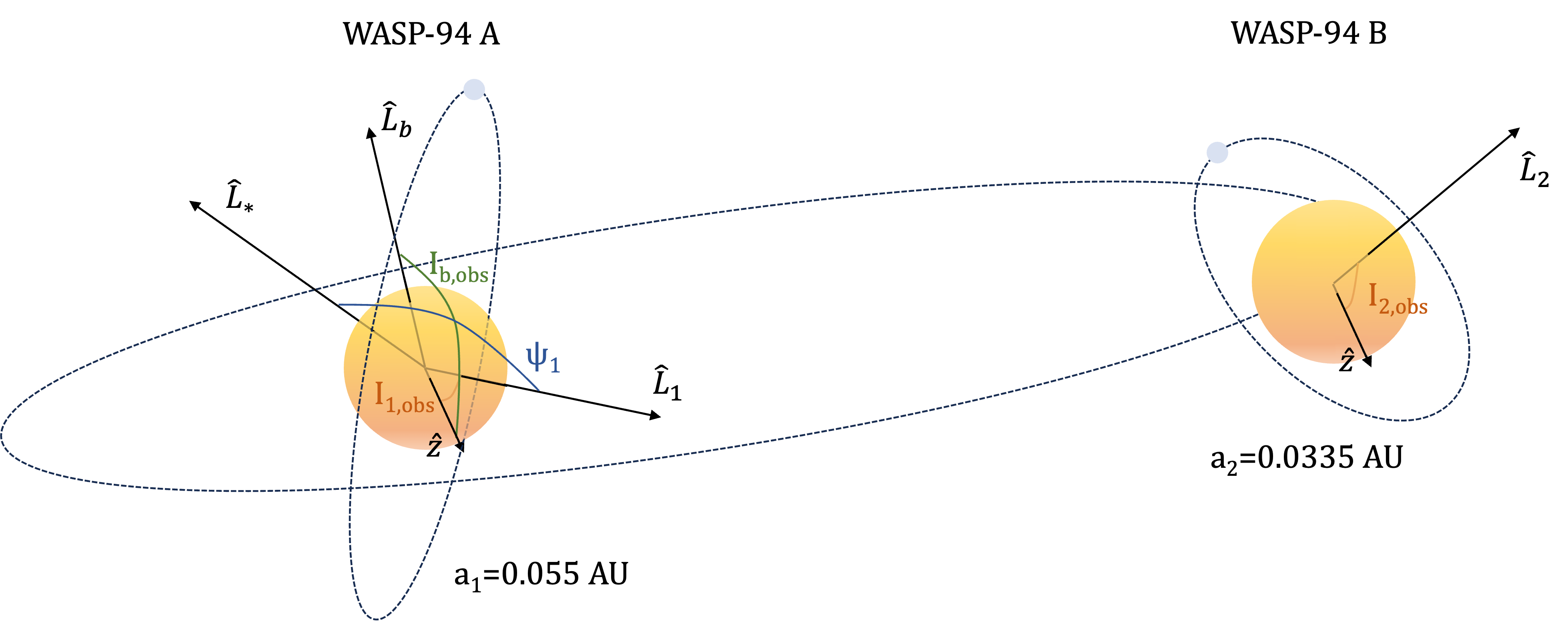}
    \caption{Schematic of the present-day WASP-94 system in the frame of WASP-94 A (orbits not to scale). The $\hat{z}$ direction points out of the page toward the observer. $\hat{L}_1$, $\hat{L}_2$, and $\hat{L}_b$ denote the orbital angular momentum vectors of WASP-94 Ab, WASP-94 Bb, and WASP-94 B, respectively. $\hat{L}_*$ denotes the stellar spin axis of WASP-94 A, and the 3D spin-orbit angle of WASP-94 Ab, which is retrograde, is represented by $\psi_1$. The inclinations of the three orbits with respect to the observer are labeled as $I_{1,\rm{obs}}$, $I_{2,\rm{obs}}$, and $I_{b,\rm{obs}}$. $I_{b,\rm{obs}}$ is unconstrained, so a fiducial value is shown. WASP-94 Ab is transiting, with $I_{1,\rm{obs}}=88.7\pm0.7^\circ$, whereas WASP-94 Bb is not, with $I_{2,\rm{obs}}\lesssim79^\circ$ or $I_{2,\rm{obs}}\gtrsim101^\circ$ \citep{neveu-vanmalle2014wasp94}.}
    \label{fig:wasp}
\end{figure*}

A schematic of the WASP-94 system is shown in Figure \ref{fig:wasp}. WASP-94 A hosts a transiting, spin-orbit misaligned ($\lambda=123\pm3^{\circ}$) hot Jupiter ($m_1=0.452^{+0.035}_{-0.032}M_{\mathrm{Jup}}$) with an orbital period of $P_1=3.95$ days, while WASP-94 B hosts a non-transiting hot Jupiter with $m_2\sin I_2=0.618^{+0.028}_{-0.029} M_{\mathrm{Jup}}$ on a $P_2=2.01$-day long orbit \citep{neveu-vanmalle2014wasp94, ahrer2024atmospheric}. The existence of this system is particularly surprising given the low ($\simeq1$\%) underlying occurrence rate of hot Jupiters measured around FGK stars \citep{howard2010california, wright2012frequency, beleznay2022exploring, Miyazaki2023sunlike}. Thus, WASP-94 raises a key question about hot Jupiter formation, as well as planet formation and evolution more generally: under what conditions may not just one, but \textit{two} such rare planets form within the same planetary system? 

There are three main pathways proposed for hot Jupiter formation: in-situ formation, disk-driven migration, and high-eccentricity tidal migration \citep{dawson2018origins}. While in-situ formation and disk-driven migration may present alternative explanations for the formation of two hot Jupiters in the WASP-94 system, in this work we consider specifically whether the high-eccentricity tidal migration pathway, and in particular the von Zeipel-Lidov-Kozai (ZLK) mechanism \citep{von_zeipel_1910, lidov1962evolution, kozai1962secular,naoz2016eccentric}, can account for the current configuration of WASP-94.

Many planet-hosting binary star systems with high initial mutual inclinations between orbits likely undergo ZLK oscillations: a mechanism that, coupled with tidal friction, has been invoked as an efficient pathway to migrate cold Jupiters inward to become hot Jupiters in hierarchical systems \citep[e.g.][]{wu2003planet, naoz2011hot, naoz2012formation}. During ZLK migration, secular interactions with a massive perturber launch a planet onto an extremely high-eccentricity orbit. Tidal forces then act to efficiently circularize and shrink the planet's orbit each periastron passage. 

Numerous studies have explored the dynamics of the ZLK effect in so-called ``2+2" system configurations, where the outer perturber is replaced by a binary system \citep[e.g.][]{pejcha_2013,hamers_2021,oconnor_2021,klein_katz_2024}. However, these previous studies generally make assumptions inappropriate for the parameter space occupied by hot-Jupiter-forming binaries -- for instance, assuming that one of the stars is much more massive than the other, or that the perturbing binary consists of two equal-mass bodies. Using \textit{N}-body simulations, \citet{liu2025formation} outlined the prospects for double hot Jupiter formation through mirrored ZLK migration, demonstrating that this mechanism can naturally produce hot Jupiters around each star in planet-hosting binary systems given sufficiently high initial mutual inclinations between the star-star and star-planet orbits. 

In this work, we conduct a detailed dynamical case study of the well-characterized WASP-94 system. Our proof-of-concept analysis demonstrates that the orbital architecture of the WASP-94 system can be naturally reproduced through mirrored ZLK migration, under the assumption that both stars in the system initially formed comparable Jovian-mass planets on wide orbits. This is the first direct, observationally-motivated case study of potential double ZLK migration in a known planetary system.

\section{Simulation Setup}
\label{sec:setup}




\begin{table*}
\centering
\begin{tabular}{lllll}
\toprule
Parameter & Definition & Initial Value & Simulation End Value & Observed Value \\\midrule
$M_1$, $M_2$ & Stellar masses ($M_\odot$) & $1.29$, $1.24$ & -- & $1.29\pm0.10$, $1.24\pm0.09$\\
$m_1$, $m_2$ & Planetary masses ($M_{\mathrm{Jup}}$) & $0.452$, $0.618$ & -- & $0.452^{+0.035}_{-0.032}$, $0.618^{+0.028}_{-0.029}$ ($m_2 \sin{I_2}$)\\
$a_1,a_2, a_b$ & Semimajor axes (au) & $7.26, 6.58, 1689$ & $0.054, 0.037, 1689$ & $0.055\pm0.001, 0.0335^{+0.0006}_{-0.0005}$, -- \\
$e_1, e_2, e_b$ & Eccentricities & $0, 0, 0.88$ & $0, 0, 0.88$ & $<0.13$, --, -- \\
$I_1,I_2,I_b$ & Inclinations ($^\circ$) & $84.4, 92.84, 0$ & $(20, 113), (87, 129), 0$ & --\\ 
$R_1,R_2$ & Stellar radii ($R_\odot$)&$1.36$, $1.35$ & -- & $1.36\pm0.13$, $1.35\pm0.12$\\
$r_1,r_2$ & Planetary radii ($R_{\mathrm{Jup}}$) &$1.665
$, $1.665$& -- & $1.665\pm0.050$, --\\
$k_{2,*},k_{2,\mathrm{Jup}}$ & Love numbers & $0.028, 0.51$ & -- & --\\
$C_*,C_{\mathrm{Jup}}$ & Gyroradii &  $0.08, 0.25$ & -- & --\\
$\Omega_*,\Omega_{\mathrm{Jup}}$ & Rotation periods (days) &  $12$, $1$ & -- & --\\
$Q_*,Q_{\mathrm{Jup}}$ & Tidal quality factors &  $10^6$, $3\times10^5$ & -- & --\\
\bottomrule
\end{tabular}
\caption{WASP-94 system parameters adopted to initialize the simulation (column 3); final values at the end of the simulation (column 4); and observed values, drawn from \citet{neveu-vanmalle2014wasp94} and \citet{Saha2024tess}, for comparison (column 5). The simulation end values for WASP-94 Bb are recorded directly before the planet is manually collided into its host star (see \S \ref{sec:results}). As WASP-94 Bb has fully circularized at this point, its orbital parameters will not be altered if we integrate the system further. Subscripts `1' refer to the WASP-94 A system, subscripts `2' to the WASP-94 B system, and subscripts `b' to the stellar binary. Subscripts `Jup' correspond to both planetary companions. All inclinations are defined relative to the binary plane as indicated by $I_b = 0^\circ$. While the planetary inclinations relative to the viewer have been observationally constrained, the binary orbit orientation is not known, preventing a direct comparison to observations. The two values in parentheses for the end-of-simulation inclinations of the planets are the lower and upper bounds of post-migration inclination precession. Structural parameters are adopted from \cite{wu2003planet}. The stellar and planetary rotation periods are adopted from \cite{naoz2016eccentric} and observational constraints for FGK stars \citep{2014ApJS..211...24M, Colman2024stellarrotation}.}
\label{tab:wasp94}
\end{table*}

\subsection{Overview}
For all simulations in this work, we used the \texttt{REBOUND}\footnote{\url{rebound.readthedocs.io}} numerical integration package \citep{rein2012rebound} with the Gragg-Bulirsch-Stoer (\texttt{BS}) integrator \citep{press02, lu2023tidespin}. General relativity and tidal forces were included using the \texttt{REBOUNDx} package \citep{tamayo2019reboundx}. General relativity was incorporated using the \texttt{gr\_full} prescription \citep{Newhall1983grfull}, which accounts for first-order post-Newtonian effects for all bodies in the system. Tidal forces were included with the \texttt{tides\_spin} prescription \citep{lu2023tidespin}, which implements self-consistent spin and dynamical evolution under the influence of tidal friction through equilibrium tides \citep{eggleton1998tidespin}. In this study we report the value of the tidal parameter $Q$, as this has been the parameter of choice adopted by the exoplanet community. However, we emphasize that the underlying tidal model used in this work is not a constant $Q$ model, but is instead a constant time lag model that is parameterized by lag time $\tau$. We convert between the two with the commonly-adopted relation $Q = (2 n \tau)^{-1}$ \citep{mardling2002scaling, lu2023tidespin}, where $n = 230$ yr$^{-1}$ is the mean motion of the planet on a $\sim10$ day orbit.

\autoref{tab:wasp94} provides a summary of the system parameters adopted in this work. We do not seek to carry out a population study -- rather, we provide an illustrative proof-of-concept for the WASP-94 system, showing that the system can be reproduced through double ZLK migration for reasonable parameter choices. Hence, we report only the initial values corresponding to the simulation that best reproduced the present-day WASP-94 configuration, out of a grid of 144 runs  -- described in \S \ref{subsection:initial_system_grid} -- that spanned a range of initial planetary semimajor axes and inclinations. All other parameters, described in \S\ref{subsection:binary_orbital_parameters},  \S\ref{subsection:planetary_orbital_parameters}, and \S\ref{subsection:additional_parameters}, are set to fiducial values that are shared across all simulations.

\subsection{Initial system grid}
\label{subsection:initial_system_grid}
In our grid, initial planetary semimajor axes ranged from 1-10 au, as giant planet occurrence rates are expected to peak near the water snow line \citep{fernandes2019hints, fulton2021california}. The planetary inclinations were drawn from a uniform distribution about $90^\circ \pm 5^\circ$ with respect to the binary's orbit, as high initial mutual inclinations have been shown to support efficient hot Jupiter formation \citep[e.g.][]{wu2003planet, naoz2012formation, liu2025formation}---that is, the maximum eccentricity reached in a ZLK cycle is maximized for near-perpendicular initial orbits.  The simulation that resulted in the closest match to WASP-94 has its initial conditions listed in \autoref{tab:wasp94}.

\vspace{1em}

\subsection{Binary orbital parameters}
\label{subsection:binary_orbital_parameters}
From \textit{Gaia} DR3, the two stars of the WASP-94 system are located at an observed separation $s=3178$ au \citep{gaia2023dr3}. The eccentricity of the binary orbit is not well-constrained, given the wide orbital separation between the two stars. However, at the population level, orbital eccentricities are observed to be systematically peaked toward higher values for wide-separation binary star systems \citep{tokovinin2015eccentricity, hwang2022wide, Hwang2022eccentricity}. 

We assume that WASP-94 A and B are observed at apoapsis, and we select an orbital eccentricity $e_b=0.88$ which corresponds to a semimajor axis of 1690 au and a periapsis of $\sim$ 200 au. This value is chosen to place the periapsis near 200 au, within which planet formation has been observed to be suppressed \citep{moe2021impact}, suggesting primordial cold Jupiter formation may be less feasible. 

Several studies \citep[e.g.][]{naoz2012formation, liu2025formation} have shown that, without a near-$90^\circ$ initial misalignment between the planetary and binary orbits, a high binary eccentricity is required to generate hot Jupiters within the age of the Universe in widely-separated ($a_* > 1000$ AU) binary systems. High binary eccentricities increase the strength of octupole-order effects, providing an alternative avenue to reach the extremely high planetary eccentricities necessary for efficient hot Jupiter generation \citep{naoz2013secular, li14chaos, weldon2024analytic}. These higher binary eccentricities further shorten the orbit flip timescale, enabling our models to reproduce WASP-94 Ab's retrograde orbit \citep{li14flip}.

Our proof-of-concept simulation includes both prerequisites (near-perpendicular mutual inclinations and high binary eccentricity) that we expect would each separately represent a sufficient criterion for hot Jupiter formation. This choice was made to minimize computation time. In principle, it should also be possible to generate the WASP-94 system with an $e_b=0$ binary orbit if the system's initial mutual inclination is near $90^\circ$, as shown by \cite{naoz2012formation}.

\subsection{Planetary radii and orbital parameters}
\label{subsection:planetary_orbital_parameters}

Both planets are assumed to originate on circular orbits that are aligned with the stellar spin axis, as is expected from strong eccentricity and inclination damping from planet-disk interactions \citep[e.g.][]{cresswell2007evolution}. The mass and radius of WASP-94 Ab are adopted from \citet{neveu-vanmalle2014wasp94} and \citet{Saha2024tess}. The true mass of WASP-94 Bb is not known, as the planet is non-transiting and only $m_2\sin I_2$ has been measured through radial velocity observations \citep{neveu-vanmalle2014wasp94}. For simplicity, we set the mass of WASP-94 Bb as the measured lower limit $0.618 M_{\mathrm{Jup}}$, noting that this is an underestimate of the true mass. The radius of WASP-94 Bb is also unknown, so we adopt $r_2=1.665R_{\mathrm{Jup}}$, matching the radius of WASP-94 Ab \citep{Saha2024tess}. 

The current orbital eccentricity of WASP-94 Bb has not been constrained, and only an upper limit has been measured for the orbital eccentricity of WASP-94 Ab. We assume that both planets' orbits have fully circularized, consistent with the assumption adopted by \cite{neveu-vanmalle2014wasp94} to fit for the planets' semimajor axes. This is a reasonable assumption for hot Jupiters, which are expected to typically reside on near-circular orbits due to their short tidal circularization timescales \citep{dawson2018origins}.

\subsection{Additional parameters}
\label{subsection:additional_parameters}
The remaining parameters in \autoref{tab:wasp94} are not constrained observationally for the WASP-94 system. Most, however, are not expected to play a significant role in our simulation results. Hence, we adopt the values used by \cite{wu2003planet}, except for the planetary and stellar rotation periods, which are drawn from \citet{naoz2016eccentric} and literature constraints for FGK stars \citep{2014ApJS..211...24M,Colman2024stellarrotation}, respectively.

The Love number $k_{2,*}$ and gyroradius $C_*$ adopted for the stars in our simulations match standard values for Sun-like stars \citep[e.g.][]{zahn1977tidal, eggleton1998tidespin}. The Love number $k_{2, \rm{Jup}}$ and gyroradius $C_{\rm Jup}$ of the planets are consistent with values inferred from measurements of Jupiter's interior \citep{iess2018jupiter,durante2020jupiter}. Our adopted $Q_*$ is another standard value for Sun-like stars and is consistent with empirical constraints derived for comparable stars \citep[e.g.][]{penev2018empirical, yee2020orbit, millholland2025empirical, Chen_2025}. The planet's tidal quality factor is expected to be more influential, and we elaborate on caveats associated with our choice of planetary tidal parameters in \S \ref{sec:tidal_params}.


\section{Results} 
\label{sec:results}
\begin{figure*}
    \centering
    \includegraphics[width=0.8\linewidth]{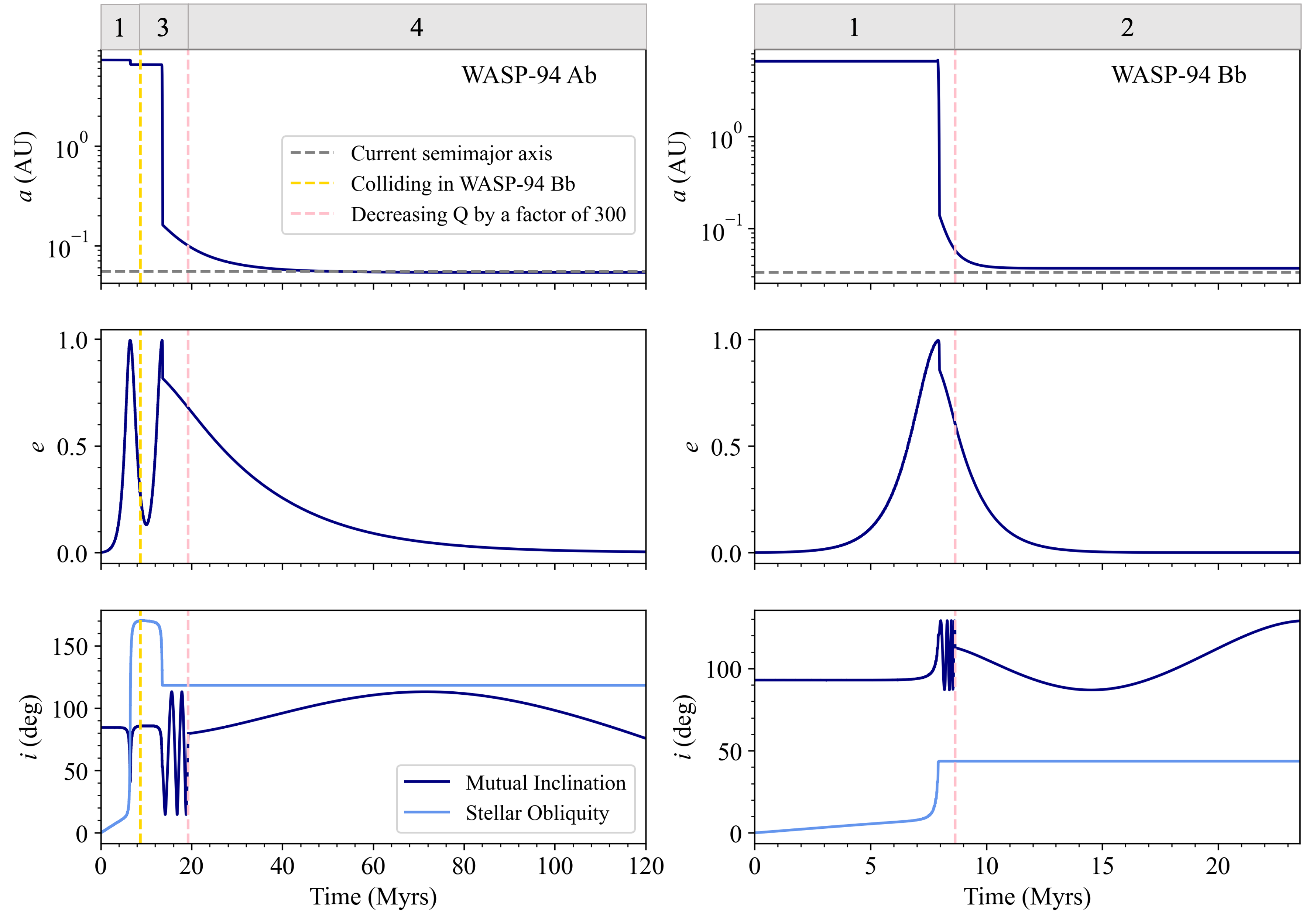}
    \caption{Semimajor axis, eccentricity, and inclination evolution of WASP-94 Ab and Bb. The gray dotted lines denote the current observed semimajor axis of each planet. The phases of the simulation are labeled at the top of each column. The yellow dotted line in the left three panels indicates the time at which WASP-94 Bb collides with its host, as viewed by WASP-94 Ab. The pink dotted lines indicate where the tidal quality factor is changed from $Q=3\times10^5$ to $Q=10^3$ for each of the two planets. We scale the time after the pink dotted lines by a factor of 300 for each planet separately. The mutual inclination evolution after the ZLK cycles end is driven by the precession of the orbit normal of the planets. This precession rate is distorted after we scale the time to account for the change in $Q$.}
    \label{fig:sim}
\end{figure*}

The results of our proof-of-concept simulation, which well reproduces the known properties of the WASP-94 system, are shown in \autoref{fig:sim}, with corresponding simulation start and end values reported in \autoref{tab:wasp94}. The computational costs required to directly run \textit{N}-body simulations of hot Jupiter systems over relevant dynamical timescales are vast: if one hot Jupiter orbit circularizes well before the other, as occurs for WASP-94 Bb in our simulation (see \autoref{fig:sim}), a direct integration would need to continue for millions of simulated years while resolving the first hot Jupiter's $<10$ day orbit. Therefore, running an end-to-end direct \textit{N}-body simulation for the entire evolution of the system would be prohibitively expensive. 

To address this problem, we have divided our simulation into four phases -- delineated by the dashed lines in \autoref{fig:sim} -- that make simplifying assumptions to enable reasonable compute times. In this section, we describe and justify each of these assumptions, arguing that they should not meaningfully impact our conclusions.

\vspace{2mm}\noindent\textit{Phase 1:} In Phase 1, the two planets simultaneously undergo ZLK oscillations. We set $Q_{\mathrm{Jup}}=3 \times 10^5$, a standard fiducial value for Jupiter-like planets, following \cite{wu2003planet}. In this phase of the simulation, the parameters are completely realistic.

\vspace{2mm}\noindent\textit{Phase 2:} The transition between Phase 1 and Phase 2 is marked where ZLK oscillations are quenched for WASP-94 Bb due to short-range forces \citep[e.g.][]{fabrycky2007shrinking, naoz2016eccentric}. At this point, we set $Q_{\mathrm{Jup}}=10^3$ for WASP-94 Bb. This effectively speeds up the eccentricity decay by a factor of 300, since the circularization timescale linearly depends on $Q$ via \citep{goldreich1966q}:

\begin{equation}
    t_{\text{circ}} = \frac{4}{63} \frac{a_{\rm{Jup}}^{13/2}}{\sqrt{GM_*^3}} Q_{\rm{Jup}} m_{\rm{Jup}} R_{\rm{Jup}}^{-5},
\end{equation}
where quantities subscripted with ``$\mathrm{Jup}$'' can be associated with either planet. We emphasize that this adjustment does not qualitatively impact the dynamical evolution of WASP-94 Bb -- only the timescale over which the planet reaches a circularized orbit. This is a common procedure in the literature for speeding up numerical simulations involving tides \citep[e.g.][]{bolmont_2015, Becker20, Lu_2024}. 

During Phase 2, the parameters of WASP-94 Ab remain unchanged. The perturbations to WASP-94 Ab are not strictly unaffected by lowering WASP-94 Bb's tidal quality factor, as from the perspective of WASP-94 Ab, WASP-94 Bb's circularization is accelerated. We argue that at this stage, WASP-94 Bb remains so close to WASP-94 B that even at the binary system's close approach of 200 au, the added perturbation from WASP-94 Bb's changing semimajor axis from $a_2=0.06$ au (at the start of Phase 2) to $a_2=0.037$ au (at the end of Phase 2) is minimal. We expect that any change in the perturbation encountered by WASP-94 Ab would, therefore, be inconsequential for the system's long-term evolution.

\vspace{2mm}\noindent\textit{Phase 3:} At this stage, the orbit of WASP-94 Bb has fully circularized. The limiting factor in computation time at this stage is the need to resolve the orbit of WASP-94 Bb, which demands an extremely small timestep when using the adaptive-timestep \texttt{BS} integrator. Therefore, for Phase 3, we elect to approximate the combined dynamical effect of WASP-94 B and WASP-94 Bb on the rest of the system as a star with an increased $J_2$ moment. 

To accomplish this, at the beginning of Phase 3 we collide WASP-94 Bb into its host star, demanding conservation of mass and angular momentum. We then artificially increase the $J_2$ moment of the star by correspondingly increasing its $k_2$ Love number, related via \citep[e.g.][]{Ragozzine2009j2k2, Millholland2019j2k2}

\begin{equation}
    \Delta k_2 = \frac{3Gm_B}{\omega^2R_B^2} \Delta J_2.
\end{equation}
The enhancement of the effective $J_2$ of a body due to a satellite is given as \citep{Tremaine1991j2, Ward2004j2, Lu2022j2}

\begin{equation}
    \Delta J_2 = \frac{1}{2}\frac{m_2}{M_2}\left(\frac{a_2}{R_2}\right)^2.
\end{equation}
This procedure increases the effective $J_2$ of WASP-94 B by a factor of $\sim 4800$. Following the same argument from Phase 2, we expect that, because WASP-94 Bb is so close to its host star, this approximation minimally affects the evolution of WASP-94 Ab.

\vspace{2mm}\noindent\textit{Phase 4:} At this stage, ZLK cycles for WASP-94 Ab have quenched and the planet's orbit begins to circularize. As in Phase 2, we decrease the tidal quality factor of WASP-94 Ab to $Q_{\mathrm{Jup}}=10^3$ and integrate until WASP-94 Ab fully circularizes.

\vspace{2mm}
The results of the full simulation, including all four phases, are shown in Figure \ref{fig:sim}. WASP-94 Bb experiences only two phases: Phase 1, which occurs before the planet's decrease in Q by a factor of 300, and Phase 2, which occurs after this decrease. The simulation end values given in Table \ref{tab:wasp94} for WASP-94 Bb are thus recorded at the end of Phase 2. Because WASP-94 Bb has fully circularized at this point, further integrating WASP-94 Bb will not alter its orbital parameters. WASP-94 Ab undergoes all four phases, but Phase 2 as experienced by WASP-94 Ab is so quick that it is not visible in the left column of Figure \ref{fig:sim}. Therefore, the three regions in the left column of Figure \ref{fig:sim} -- divided by the two vertical dashed lines -- effectively correspond to Phase 1, Phase 3, and Phase 4.

To recover realistic timescales for \autoref{fig:sim}, for each planet we scaled the time after the tidal quality factor change by a factor of 300 in accordance with the change in eccentricity decay timescale. We find that both WASP-94 Ab and Bb successfully evolve into hot Jupiters in under 60 Myr. After the planets' orbits are fully circularized, their final semimajor axes $a_{1, \rm{sim}} = 0.054$ au and $a_{2, \rm{sim}}=0.037$ au closely match the observed values of $a_{1,\rm{obs}}=0.055\pm0.001$ au and $a_{2, \rm{obs}}=0.0335^{+0.0006}_{-0.0005}$ au. 

Relative to the viewer, WASP-94 Ab has an observed orbital inclination $88.7\pm0.7^\circ$, and  WASP-94 Bb has an orbital inclination constrained to $\lesssim79^\circ$ or $\gtrsim101^\circ$, indicating that the two planets' orbits are mutually inclined \citep{neveu-vanmalle2014wasp94}. However, the inclination of the binary orbit is not well-constrained, such that we cannot directly compare these observations with the \autoref{tab:wasp94} values that are provided in the system's inertial frame. The final mutual inclinations of the planets in our simulation undergo oscillations, with $I_{1,\rm{sim}}$ oscillating between $20^\circ$ and $113^\circ$ and $I_{2,\rm{sim}}$ oscillating between $87^\circ$ and $129^\circ$. These ranges are consistent with the two planets being mutually inclined with respect to each other, and are thus in agreement with observations.

The bottom left panel of Figure \ref{fig:sim} shows that the final spin-orbit orientation of our simulated WASP-94 A b is retrograde, in agreement with the observed $\lambda=123\pm3^{\circ}$. Because only the projected spin-orbit angle $\lambda$ has been measured, rather than the true spin-orbit angle $\psi$ that is probed in our simulations, we cannot conduct a direct comparison between these values.

\section{Discussion}

\subsection{Tidal Parameters}
\label{sec:tidal_params}
The values of the tidal parameters $Q_*$ and $Q_{\rm Jup}$ are highly uncertain. Jupiter's $Q$ value is constrained to $\approx10^5$ from Io's secular migration \citep{lainey2009strong}, but exoplanet $Q$ values are far less certain \citep[e.g.][]{morley_2017, puranam_chaotic}. Similarly, measurements of the Love number $k_2$ in exoplanets are rare and highly uncertain \citep[e.g.][]{buhler2016hatp13,bouma2019wasp4,csizmadia2019wasp18, wahl2021tidal}. These parameters primarily affect the timescale of evolution; as seen from \autoref{fig:sim}, WASP-94 evolves to the present-day configuration comfortably within the age of the system.

Another substantial simplification is the use of the equilibrium tide model, which includes assumptions that break down in the regime of highly eccentric orbits. Tidal models that account for the planet's interior structure in a more nuanced fashion, such as dynamical tides \citep[e.g.][]{mardling_1995, lai_dynamical,liveoack2025dynamical} and inertial waves \citep{dewberry2024tidal,dewberry2025inertial}, are likely more accurate for models of ZLK migration. However, given the many degrees of freedom necessarily introduced by more detailed tidal models, we elected to use the simpler equilibrium tide framework to capture the overarching system behavior while maximizing physical intuition. We note that the first-order impact of dynamical tides would be increased energy dissipation at pericenter approach \citep{vick_2019,yang2025hatp7}; as such, we may expect more rapid evolution toward the present-day WASP-94 system configuration if dynamical tides are taken into account.

\subsection{Primordial Misalignment}
Dissipative precession at the circumstellar disk stage, as examined in detail for exoplanet systems in \citet{gerbig2024aligning}, has the potential to push exoplanet-hosting binary star systems toward low mutual orbital inclinations, such that ZLK oscillations would be suppressed. However, previous studies have shown that such trends are most prevalent for stellar binary separations $s\lesssim700-800$ au \citep{christian2022possible, dupuy2022orbital, lester2023visual, gerbig2024aligning} and that the overarching trend toward alignment is weaker for hot-Jupiter-hosting systems \citep{christian2025wide}. 

This weaker alignment for hot-Jupiter-hosting binary systems, which instead show evidence for ZLK oscillations at the population level, is particularly marked in binaries that include a hot stellar host with $T_{\rm eff}\gtrsim6100$ K \citep{rice2024orbital} -- as is the case for the WASP-94 system, where WASP-94 A is measured at $T_{\rm eff}\approx6170$ K \citep{neveu-vanmalle2014wasp94, teske2016curious, bonomo2017gaps, stassun2017accurate}. Therefore, while early alignment from dissipative precession will prevent a subset of systems from undergoing double ZLK oscillations, WASP-94 falls within the parameter space of binary systems that are not as strongly affected by this process.

Theoretical expectations from protoplanetary disk formation models and empirical evidence from ALMA observations of young stellar systems further justify the choice of primordially misaligned orbits in this work. Widely-separated binary systems that formed via turbulent fragmentation are expected to be randomly oriented, with uncorrelated angular momentum vectors \citep{offner2016turbulent, Bate2018}. Consistent with this picture, numerous individual wide binary systems have been observed with misalignments \citep{Jensen2014misaligned, Akeson2014disks, offner2023origin}. At a population level, while compact binaries are preferentially aligned, the distribution of wide binaries has been found to be statistically distinct \citep[e.g.][]{Reynolds2024diskorientations}.

\subsection{Binary Orbital Eccentricity}
The fiducial orbital eccentricity selected for the WASP-94 stellar binary within this work is high, at $e_b=0.88$, to ensure that ZLK oscillations operate efficiently. High-eccentricity stellar binaries are common at wide orbital separations $>10^3$ au, where a superthermal distribution $f(e)\propto e^{\alpha}$, with $\alpha>1$, describes the eccentricity distribution \citep{Hwang2022eccentricity}. Given the prevalence of highly eccentric stellar binaries at wide separations, and considering the separation $s=3178$ au between the WASP-94 host stars, our selected eccentricity in this work falls within a reasonable expected range of values.

While the true binary orbital eccentricity may differ from $e_b=0.88$, a high value for $e_b$ improves the efficiency of double ZLK migration as a formation channel for the two WASP-94 hot Jupiters, as addressed in \S \ref{sec:setup}. If the binary orbit is ultimately constrained to lie closer to circular, this would either suggest that the two hot Jupiters formed largely independently -- through channels such as disk migration \citep{lin1996migration}, in situ formation \citep{batygin2016situ}, or planet-planet driven high-eccentricity migration \citep[e.g.][]{wu2011secular,beauge2012multiple, petrovich_2015}; that ZLK migration is far more efficient at forming hot Jupiters than our model suggests (e.g., if dynamical tides are highly significant); or that the initial mutual inclination between the binary and planetary orbits was near-perpendicular. We also note that our choice of $e_b=0.88$ places the periastron distance between the two stars in our system at $q=200$ au, which is wide enough that we do not expect substantive disk truncation that may inhibit cold Jupiter formation \citep{moe2021impact}. Substantially higher orbital eccentricities may limit the formation of wide-orbiting giant planets. 

\subsection{Opportunities Afforded by the Gaia Mission}


The \textit{Gaia} space mission \citep{gaia2016} has completed its 10+ years of observations and will ultimately provide time-series astrometric data over this baseline, leading to the projected discovery of 120,000$\pm$22,000 giant planets at typically 1-5 au orbital separations \citep{lammers2025exoplanet}. If the process of double ZLK migration is, indeed, common -- which will depend on the joint occurrence of cold Jovian planets and high-eccentricity close- to moderate-separation binary star systems -- we may expect cold Jovians undergoing ZLK oscillations to be discovered around one or both stars in binary systems. For systems with insufficiently high mutual inclinations to induce ZLK oscillations, the rate of cold Jovians around both stars in stellar binary systems will also offer insights into the occurrence of systems with the initial conditions necessary to invoke double ZLK migration. 

Crucially, \textit{Gaia} will provide mutual orbital inclinations and thus full 3D architectures of these systems -- a remaining degree of uncertainty in the present study. The orbital solutions provided by \textit{Gaia} will present the first population of long-period planets in binary star systems with which we can statistically probe the mechanisms behind double hot Jupiter formation, as well as the ZLK mechanism more generally. 
Follow-up of candidate planets orbiting highly eccentric, moderate-separation stellar binaries with periastron approaches $q\gtrsim200$ au would offer critical insights into the prevalence of systems undergoing double hot Jupiter formation.

\section{Conclusion}
\label{sec:conclusions}
It is striking that a pair of stars as remote as $3000$ au can force a pair of planets to migrate inwards. In this work, we have leveraged numerical simulations to demonstrate that the configuration of the WASP-94 double hot Jupiter system can be naturally generated through mirrored ZLK migration.

By experimenting with different sets of initial planetary semimajor axes and mutual inclinations, we have identified a configuration that can well reproduce the observed parameters of the WASP-94 system. In particular, the final semimajor axes and orbital eccentricities of the two planets match the observationally constrained values, and WASP-94 Ab becomes retrograde at the end of the simulation, in agreement with observations.

WASP-94 is one potential example of the double ZLK mechanism that may play a role in many exoplanet-hosting binary star systems. Forthcoming data releases from the \textit{Gaia} mission will offer new insights into the prevalence of this mechanism, as well as the extent to which ZLK dynamics shape the present-day 3D architectures of planetary systems. By combining \textit{Gaia} data with follow-up radial velocity observations, it will be possible to, for the first time, reconstruct the full 3D architecture of cold Jovian planets' orbits around both stars in binary systems. In this way, \textit{Gaia} offers a concrete and profound avenue forward to understand the prevalence of double ZLK migration, and to conduct comparative demographics in exoplanet systems more generally.

\section*{Acknowledgements}
The first sentence of \S \ref{sec:conclusions} is an intentional homage to \cite{wu2003planet}. We are grateful for this paper and others that have laid the bedrock for understanding how hot Jupiters may form through ZLK migration, setting the foundation for this work.

T.L. and M.R. acknowledge support from Heising-Simons Foundation Grant \#2021-2802. T.L. is supported by a Flatiron Research Fellowship at the Flatiron Institute, a division of the Simons Foundation. M.R. acknowledges support from Heising-Simons Foundation Grant \#2023-4478, National Geographic grant \#EC-115062R-24, and NASA Exoplanets Research Program NNH23ZDA001N-XRP (grant No. 80NSSC24K0153). This work has benefited from the use of the \textit{Grace} computing cluster at the Yale Center for Research Computing (YCRC). 

\software{\texttt{REBOUND} \citep{rein2012rebound}, \texttt{REBOUNDx} \citep{tamayo2019reboundx}, \texttt{matplotlib} \citep{hunter2007matplotlib}, \texttt{numpy} \citep{oliphant2006guide, walt2011numpy, harris2020array}, \texttt{pandas} \citep{mckinney2010data}, \texttt{scipy} \citep{virtanen2020scipy}}

\bibliography{bibliography_dhj.bib}
\end{CJK*}
\end{document}